# Fast implementation of synchrosqueezing transform based on downsampling for large-scale vibration signal analysis

Dong He,　Hongrui Cao[*]

*State Key Laboratory for Manufacturing Systems Engineering, Xi'an Jiaotong University, Xi'an, Shaanxi 710049, China*

## Abstract

Synchrosqueezing transform (SST) is a useful tool for vibration signal analysis due to its high time-frequency (TF) concentration and reconstruction properties. However, existing SST requires much processing time for large-scale data. In this paper, some fast implementation methods of SST based on downsampled short-time Fourier transform (STFT) are proposed. By controlling the downsampling factor both in time and frequency, combined with the proposed selective reassignment and frequency subdivision scheme, one can keep a balance between efficiency and accuracy according to practical needs. Moreover, the reconstruction property is available, accomplished by an approximate but direct inverse formula under downsampling. The effects of parameters on the concentration, computing efficiency, and reconstruction accuracy are also investigated quantitatively, followed by a mathematic model of reassignment behavior with decimate factors. Experimental results on an aero-engine and a spindle show that the fast implementation of SST can effectively characterize the non-stationary characteristics of the large-scale vibration signal to reveal the mechanism of mechanical systems.

Corresponding author. Tel.: +86 29 82663689; fax: +86 29 82663689.

*E-mail address*: chr@mail.xjtu.edu.cn (Hongrui Cao).

# 1. Introduction

Non-stationary signal feature extraction is one of the key parts in machinery fault diagnosis. In order to analyze non-stationary signals generated by machinery operating under non-stationary working conditions, synchrosqueezing transform (SST) [1-4] is widely applied [5-8] and extended to many variations [9-16]. This is because SST can improve the concentration of the time-frequency (TF) representation (TFR) of non-stationary signals composed of multiple components, more importantly, without losing reconstruction properties [17]. However, the standard implementation of SST is rather time-consuming for large-scale data because the time-frequency matrices are large; more matrices are calculated except short-time Fourier transform (STFT); the reassignment operation is implemented in full frequency band. Thus, how to perform SST efficiently for large-scale data, making it more capable of practical engineering, is worth studying.

The SST can be regarded as the combination of conventional linear TF transform and reassignment operation along the frequency domain. Therefore, one can accelerate SST by downsampling the TF transform and improving the reassignment strategy, and in the meantime, should consider the reconstruction property. In this regard, the fast implementation of synchrosqueezing transform could be also termed invertible downsampled synchrosqueezing transform. In this paper, the STFT-based synchrosqueezing transform and its relevant issues are studied, but they can be extended to a wavelet-based version.

The standard SST and its inverse require that the decimation factor of the time axis equals 1, which results in large TF matrices. Fortunately, reducing the size of the TF matrices has been well studied, which is accomplished by discretizing or downsampling the classical STFT [18, 19] by controlling the downsampling factor (or decimate factor) in the time domain and frequency domain. The decimate factor in the time domain is equivalent to the hop size of the window $H_t$. In the frequency domain, the number of points of discrete Fourier transform (DFT) $N_f$ controls the discretization. For the standard full sampled STFT, the $N_f$ is usually equal to the whole length of the collected signal. Thus, for consistency, the downsampling factor in frequency $H_f$ is defined as the ratio of $N$ to $N_f$. In order to reduce the computation time of STFT without causing distortion, we can appropriately increase the factor $H_t$ and $H_f$, i.e. reduce the time-frequency sampling density. However, this downsampling idea needs to be incorporated in the SST framework and the effect of the decimate factor on the quality of TFR needs to be investigated.

Some acceleration implementations of SST have been made based on discrete STFT or more general TFRs. Fenet et al. [20] formulated a general framework for discrete-time reassignment method (RM) and SST and applied it to constant-Q transform. However, the hop size of the window and the invertibility of the framework need to be discussed. Considering the hop size and invertibility, the work reported by Holighaus et al. [21] extended the RM and SST to general time-frequency filters with arbitrary decimation factors by the frame property. Another study that makes SST suitable for real-time computation was accomplished by Fourer et al. [22]. They proposed a recursive STFT-based SST and then extended it to second order for instantaneous frequency (IF) estimation [23]. Their method was implemented recursively by filters banks in a small frequency range instead of the whole frequency band, so the computational cost can be reduced. Nevertheless, the number of columns of TF matrix is still equal to the number of samples so it requires large storage.

Another issue of standard SST is that the existing reassignment operation is performed on full frequency bins. This is time-consuming, especially $N_f$ is large. In some practical conditions, we tend to pay attention to the TF information in a certain frequency band. In this case, the frequency band of interest can be isolated according to a priori, so relocating the TF coefficients in this frequency band is another strategy to boost SST. The reduction of the frequency band in [22] can be used for boosting reassignment operation and mode retrieval. A similar idea can be found in [7, 24], where the called zoom SST was used to tighten the frequency interval in a preselected frequency/scale band but not for acceleration.

Mode reconstruction is one of the most significant properties of SST and is desired by many application fields. The existing inverse SST is implemented by accumulating the TF coefficients on each column [4, 22, 25]. Nevertheless, after downsampling, the number of columns of the SST matrix is not equal to the number of signal points anymore, so the signal cannot be recovered correctly via the standard inverse formula. Thus, the reconstruction of SST should be modified for downsampled SST. An alternative solution is seeking help from classical downsampled STFT synthesis where the inverse FFT and overlap-add are employed. Meignen et al. [26] studied the influence of the downsampling in the time domain on mode retrieval from downsampled STFT. On the contrary, for SST, the effect of downsampling in the frequency domain is more worth studying as it is more significant than that in the time domain. When applying STFT synthesis to SST, the TF coefficients have been relocated so they are inversely reassigned via a map before STFT synthesis [21]. To separate the mode of interest, one can isolate its ridge on RM/SST TFR and then map the TF points outside this isolated region to the TF points of STFT using the inverse reassignment map. A filtered invertible STFT with downsampled factor is thus obtained indirectly.

In this paper, some implementation issues of SST based on downsampled STFT is studied. Moreover, the selective reassignment and frequency subdivision are combined not only to further accelerate SST but also to obtain more degree of freedom for adjusting frequency intervals according to needs. In terms of invertibility, instead of inverse mapping [21] before STFT synthesis, we are exploring the possibility of using an approximate but more direct way to reconstruct components from downsampled SST. This means that the inverse formula of standard SST is modified, making downsampled SST nearly have the same properties as classical SST. A mathematic model of reassignment behavior with decimate factors is also presented for the tradeoff between efficiency and concentration. The effects of parameters (decimate factor, selective reassignment and bandwidth) on the concentration, computing efficiency, and reconstruction accuracy are also studied quantitatively.

The remainder of this paper is organized as follows. In Section 2, the downsampled STFT and SST are reviewed, followed by a modified approximate inverse formula, selective reassignment, and frequency subdivision. In Section 3, the parameter selection is studied by a numerical model. A mathematic model of reassignment behavior with decimate factors is presented in Section 4. Then the proposed method is tested on an aero-engine and time-varying chatter feature extraction in Section 5. Finally, conclusions are drawn in Section 6.

## 2. Downsampled STFT-based synchrosqueezing transform

As stated above, the downsampled SST is based on STFT, so continuous and downsampled STFT are reviewed ahead in Section 2.1, followed by its downsampled SST in Section 2.2. The proposed reconstruction formula and reassignment scheme are introduced in Section 2.3 and 2.4, respectively.

### 2.1 Downsamped discrete STFT

The STFT of $x(t)$ using a sliding analysis window $g(t)$ is expressed as

$$S_x^g(u,\xi) = \int x(t)g(t-u)e^{-i\xi(t-u)}\,dt \tag{1}$$

where $u$ and $\xi$ are time delay and frequency, respectively. In this paper, we use Gaussian window $g(t) = (\pi\sigma^2)^{-1/4}e^{-t^2/2\sigma^2}$.

On the other hand, the inverse STFT lays the foundation of the reconstruction property of SST:

$$y(t) = \frac{1}{2\pi}\iint S_x^g(u,\xi)\gamma(t-u)e^{i\xi(t-u)}\,dud\xi \tag{2}$$

where $\gamma(t-u)$ is called synthesis window. Substitute Eq. (1) to Eq. (2) and use the property of delta function, we have

$$y(t) = x(t)\int g^*(t-u)\gamma(t-u)du = x(t)\int g^*(u)\gamma(u)du \tag{3}$$

Substitute Eq (2) to Eq. (3), we obtain

$$x(t) = \frac{y(t)}{\int g^*(u)\gamma(u)du} = \frac{1}{2\pi\int g^*(u)\gamma(u)du}\iint S_x^g(u,\xi)\gamma(t-u)e^{i\xi(t-u)}\,dud\xi \tag{4}$$

The synthesis window $\gamma(u)$ can have any form. If $\gamma(u)=\delta(u)$, $\int g^*(u)\delta(u)du = g(0)$. The reconstruction formula can be simplified as

$$x(t) = \frac{1}{2\pi g(0)}\iint S_x^g(u,\xi)\delta(t-u)e^{i\xi(t-u)}\,dud\xi = \frac{1}{2\pi g(0)}\int S_x^g(t,\xi)d\xi \tag{5}$$

which is used to derive the standard inverse SST where the signal is reconstructed by accumulating STFT along the frequency axis because each column of the TF matrix is corresponding to a time instant.

If $\gamma(u)=g(u)$, the synthesis window is not $\delta(t)$ any more and the reconstruction formula cannot be simplified as Eq. (5). According to Eq. (2) we have

$$x(t) = \int \left[ \frac{1}{2\pi} \int \tilde{S}_x^g(u,\xi) e^{i\xi t} d\xi \right] g(t-u) du = \int \tilde{s}(u,t) g(t-u) du \quad \text{s.t.} \int |g(t)|^2 = 1 \quad (6)$$

where $\tilde{S}_x^g(u,\xi) = S_x^g(u,\xi) e^{-i\xi u}$. Eq. (6) is the basis of the reconstruction formula of downsampled SST. It is noteworthy that the window $g(t)$ should be normalized by $\int |g(t)|^2 = 1$.

In the downsampling of discrete-form STFT, the analysis window $g[n]$ with length $L$ slides on the signal $x[n]$ and moves $H_t$ points each time.

$$S_x^g[m,k] = \sum_{n \in \mathbb{Z}} x[n] g[n - mH_t] e^{-i2\pi k(n - mH_t)/N_f} \quad (7)$$

where $m$ and $k$ are the index of time and frequency sequence, respectively. The downsampling rate increases with $H_t$ and $H_f$, which means higher efficiency but lower accuracy. Considering the discrete window $g[n]$ remains identical for each slide and it supports on $[-M, M]$, $M=(L-1)/2$ and the support of $g[n]$ is $[0, L-1]$, the convolution can be simplified as a practical implementation of DFT

$$S_x^g[m,k] = \sum_{n=0}^{L-1} x[n + mH_t - M] g[n - M] e^{-i2\pi k(n-M)/N_f} \quad (8)$$

In order to refine the frequency axis and remain efficiency simultaneously, set the number of points of DFT to $N_f$ ($L := 2M+1 \leq N_f \leq N$)

$$S_x^g[m,k] = e^{i2\pi kM/N_f} \cdot \sum_{n=0}^{N_f-1} x[n + mH_t - M] g[n - M] e^{-i2\pi kn/N_f} \quad (9)$$

One can further accelerate DFT in Eq. (9) by setting $N_f$ to the power of 2 and using FFT algorithm.

For STFT synthesis with decimate factors, Eq. (6) is used instead of Eq. (5) for generating more stable result as

$$x[n] = \sum_{m \in \mathbb{N}} \sum_{k=0}^{N_f - 1} S_x^g[m,k] g[n - mH_t] e^{i2\pi k(n - mH_t)/N_f}$$

$$x[n + mH_t - M] = \sum_{m \in \mathbb{N}} g[n] \sum_{k=0}^{N_f - 1} \tilde{S}_x^g[m,k] e^{i2\pi kn/N_f} \tag{10}$$

where $\tilde{S}_x^g[m,k] = S_x^g[m,k] e^{-i2\pi kM/N_f}$ and $n \in [-M, M] \cap \mathbb{Z}$. This means that the signal can be recovered through the multiplication between $S_x^g[m,k]$ and the fixed phase term $e^{-i2\pi kM/N_f}$, followed by $N_f$-point inverse DFT. Finally, an overlap-add method is employed to calculate $x[n + mH_t - M]$ at discrete time instants

$$x[n + mH_t - M] = \sum_{m \in \mathbb{N}} g[n] \cdot \tilde{s}[m,n], \quad n \in [0, L-1] \cap \mathbb{Z} \tag{11}$$

where $\tilde{s}[m,n] = F_{N_f}^{-1} \{\tilde{S}_x^g[m,k]\}$. We should consider that $\sum_{m \in \mathbb{N}} |g[n - mH]|^2 = 1$ for normalization as well.

2.2 *Downsampled Synchrosqueezing transform*

The downsampled STFT-based SST is performed by three steps. The first step is the calculation of two STFTs $S_x^g[m,k]$ and $S_x^{g'}[m,k]$ respect to $g[n]$ and $g'[n]$ by Eq. (9), respectively, where $g'[n]$ is sampled from $g'(t) = -(\pi\sigma^2)^{-1/4} e^{-t^2/2\sigma^2} \cdot t / \sigma^2$.

In a discrete TFR, the angular frequency interval can be expressed as

$$\Delta \xi = 2\pi \Delta f = 2\pi f_s / N_f \tag{12}$$

and $\xi[k] = k\Delta\xi$ is the discrete angular frequency sequence, $k=0,1…, N$-1. Then the discrete IF estimator $\hat{\omega}_x[m,k]$ for the discrete signal $x[n]$ is calculated by

$$\hat{\omega}_x[m,k] = \left| \xi[k] - \Im \left\{ \frac{S_x^{g'}[m,k]}{S_x^g[m,k]} \right\} \right|, \quad S_x^g[m,k] > \gamma \tag{13}$$

where $\gamma$ and $\Im[\cdot]$ denotes the threshold and the imaginary part of a complex number, respectively.

The final step is to reassign the TF coefficients from the computational position to the center of gravity in the frequency direction, which is realized by summing different contributions from their original position to where the IF estimator points.

$$T_x[m,l] = \sum_{k:|\hat{\omega}_x[m,k]-\xi[l]|\leq\Delta\xi/2} S_x^g[n,k] \tag{14}$$

Let $l \in \mathcal{M}_q[m]$ be the indices of the zone around the $q$th component $x_q$. This zone is constructed by extracting TF ridges representing IFs and then offsetting the IFs in both directions to form a banded area. For the standard SST in which time decimate factor equals 1, the mode can be retrieved like Eq. (5) by adding up TF coefficients in TF band zone $\mathcal{M}_q[m]$ for each column of the TF matrix $T_x[m,l]$

$$x_q[m] = \frac{1}{2\pi g(0)} \Re\left\{\sum_{l \in \mathcal{M}_q[m]} T_x[m,l]\right\} \tag{15}$$

where $\Re[\cdot]$ denotes the real part of a complex number and the number of $m$ is equal to the number of original signal, so $x[m]$ can reflect the desired signal components. However, this method is invalid for downsampled SST since the length of the accumulated result does not match the original signal.

2.3 *Reconstruction from downsampled synchrosqueezing transform*

In this subsection, an approximate but simple inverse formula based on downsampled STFT is designed for downsampled SST. The inverse SST in continuous form is written as [4]

$$T_x(u,\omega) = \int_{\Xi_{x,S(u)}} S_x^g(u,\xi)\delta(\omega-\hat{\omega}_x(u,\xi))\,d\xi \tag{16}$$

It is assumed that a slow-varying mono-component signal $x = x_k$ is analyzed, so the $\hat{\omega}_x(u,\xi)$ can be approximated by true IF $f_k(u)$, so substitute Eq. (1) to Eq. (16), we have the approximated SST TFR as

$$T_x(u,\omega) \approx 2\pi g(0)x(u)\delta(\omega-f_k(u)) \tag{17}$$

which shows that the SST for slow-varying components is almost a well-concentrated delta function on TF plane, centered around the true IF.

For comparison, the approximation of STFT in continuous form of signal $x_k(t)$ is [4]

$$S_{x_k}^g(u,\xi) \approx x_k(u)\hat{g}(\xi-f_k(u)) \tag{18}$$

where $\hat{g}$ denotes the Gaussian window defined in the frequency domain. For a fixed time, this shows STFT has the bell-shaped frequency diffusion centered around the ridge. Then the signal can be recovered via inverse Fourier transform (FT) as Eq. (10). However, for downsampled SST, the TF coefficients have been moved along frequency direction during the reassignment step. In spite of this, they are still approximately squeezed to a delta function from the bell shape of STFT if the slow-varying assumption is satisfied. It can be anticipated there will be reconstruction errors in practice and the error increases with the deviation from the slow-varying assumption, i.e., the error increases with $|f_k'(u)|$.

Thus, under this assumption, it is attempted to apply inverse FT to downsampled SST like Eq. (6)

$$y(t) = \frac{1}{2\pi} \iint T_x(u,\omega) g(t-u) e^{i\omega(t-u)} \, d\omega du \tag{19}$$

Substituting Eq. (17) into Eq. (19) we have

$$\begin{aligned} y(t) &= g(0) \iint x(u) \delta(\omega - f_k(u)) g(t-u) e^{i\omega(t-u)} \, d\omega du \\ &= \int x(u) [g(0) g(t-u)] e^{if_k(u)(t-u)} \, du \end{aligned} \tag{20}$$

On the other hand, for STFT, the reconstruction formula can be approximated as

$$\begin{aligned} y(t) &= \frac{1}{2\pi} \iint x(u) \hat{g}(\xi - f_k(u)) g(t-u) e^{i\xi(t-u)} d\xi du \\ &= \int x(u) [g(t-u)]^2 e^{if_k(u)(t-u)} du \end{aligned} \tag{21}$$

Comparing Eq. (20) and Eq. (21), for reconstruction from STFT in Eq. (21), the Gaussian window should have unit energy, i.e. $\int |g(t)|^2 = 1$. However, if using this normalized Gaussian window $g$ directly, the reconstruction condition for SST in Eq. (20) does not hold since $\int |g(0)g(t-u)| dt \neq 1$. In fact, the reconstruction of Eq. (20) amounts to using $g(t)$ and $g(0)$ as analysis and synthesis window, respectively, so we have

$$\int |g(0)g(t)| dt = \int \left[ (\pi\sigma^2)^{-1/4} \right]^2 e^{-t^2/2\sigma^2} \, dt = \sqrt{2} \tag{22}$$

which means that transferring the predefined normalized Gaussian window $g$ to SST frame is equivalent to using virtual window $h$ in STFT frame to process the signal, so

we have

$$g(0)g(t)=[h(t)]^2 \quad (23)$$

and the exact form of this equivalent window $h$ is

$$\begin{aligned} h(t) &= (\pi\sigma^2)^{-1/4}\,e^{-t^2/4\sigma^2} \\ C &= \int |h(t)|^2 = \sqrt{2} \end{aligned} \quad (24)$$

This constant can be understood as the correction term if using the normalized Gaussian window $g$ to process SST directly.

Now we present the reconstruction of downsampled SST in discrete form. After multiplying $T_x[m,l]$ by the phase term $e^{-i2\pi kM/N_f}$, perform $N_f$-point inverse DFT to each column of SST matrix and get the real part

$$\tilde{r}[m,n] = \Re\left(F_{N_f}^{-1}\left\{T_x[m,l]e^{-i2\pi kM/N_f}\right\}\right) \quad (25)$$

Then the same overlap-add formula as Eq. (11) and normalization constant in Eq. (24) are applied to recover the signal

$$x[n+mH_t-M] = \frac{1}{\sqrt{2}}\sum_{m\in\mathbb{N}}\tilde{r}[m,n]\cdot g[n],\ n\in[0,L-1]\cap\mathbb{Z},\ \sum_{m\in\mathbb{N}}|g[n-mH]|^2 = 1 \quad (26)$$

In order to separate different components from multicomponent signals, time-frequency filtering is applied. Just like full sampled SST, the reconstruction under the downsampling scheme can be realized by setting selected time-varying frequency band to separate each component $x_q$.

2.4 *Selective reassignment and frequency subdivision*

It is desirable that the reassignment process stated above can be further accelerated. In this subsection, the selective reassignment combined with frequency subdivision is proposed to reduce the computational complexity without losing too much TF characterizing ability. Looking back at the rearrangement operation, all the time-frequency points need to be tested in this process. Therefore, it is time-consuming when the TF matrix is very large. For practical engineering applications, we tend to focus on TF information in a certain frequency range. In this case, according to a prior of the signal frequency structure, the frequency band of interest can be artificially selected.

The reassignment is then performed in the selected frequency band, thereby reducing the competitional burden. This idea is called selective reassignment.

Assuming the selected frequency band range is [$f_1$, $f_2$], so the frequency sequence after the selection becomes $f_1 < \xi^*[k] < f_2$. Therefore, the selective reassignment operation can be mathematically described as

$$T_x[m,l] = \sum_{k:\left|\hat{\omega}_x[m,k]-\xi^*[l]\right|\leq \Delta\xi/2} S_x^g[m,k] \tag{27}$$

The only difference between Eq. (14) and Eq. (27) is that $\xi[l]$ is replaced by $\xi^*[l] = 2\pi[f_1 + l(f_2 - f_1)/N_r]$ given the number of frequency bins between $f_1$ and $f_2$ is $N_r$ and $\Delta\xi = 2\pi(f_2 - f_1)/N_r = 2\pi H_f f_s / N$. Obviously, selective reassignment reduces the size of the TF matrix waiting for reassignment to further accelerate SST. Also, one can truncate $S_x^g$ and $S_x^{g'}$ according to the selected frequency band before calculating $\hat{\omega}_x$ to save more space. Another benefit of selective reassignment is the ridge detection can be accelerated as well, thanks to the reduced size of the SST matrix.

In practical discrete conditions, the reassignment is to find the indices on frequency bins corresponding to the IF estimator through rounding operation. The reassignment accuracy depends on round-off error, i.e. frequency interval for reassignment. Thus, to relocate the TF coefficients more accurately, the frequency interval should be as small as possible. It should be noted that the aforementioned selective reassignment itself cannot narrow the frequency interval; it just narrows down the *global reassignment range $f_1$~ $f_2$* for faster speed. Instead, one can improve the reassignment accuracy and reduce the round-off error by subdividing the frequency intervals, like zoom SST [7, 24]. However, this may increase the computational burden in turn. Fortunately, after selective reassignment, the frequency band has changed to a small subset of the whole frequency axis. Therefore, the subdivision is performed only on this selected frequency band with factor $z$ and the effect of frequency subdivision on the overall efficiency of SST is limited. This operation is called frequency subdivision, which can be formulated as

$$T_x[m,l] = \sum_{k:|\hat{\omega}_x[m,k]-\xi^*[l]|\leq \Delta\xi^*/2} S_x^g[m,k] \qquad (28)$$

where $\Delta\xi^* = \Delta\xi/z = 2\pi H_f f_s/zN$ and

$$\xi^*[l] = 2\pi[f_1 + \frac{l}{zN_r}(f_2 - f_1)] \qquad (29)$$

For pure selective reassignment without frequency subdivision, $z = 1$ and Eq. (28) is degenerated to Eq. (27).

2.5 *Computational complexity analysis*

This subsection analyzes the computational complexity quantificationally. The number of floating-point operations times for $S_x^g[m,k]$ and $S_x^{g'}[m,k]$ is about $(L+N_f \log_2 N_f)N/H_t + NN_f/H_t$, where the number of columns is $N/H_t$. When calculating $\hat{\omega}_x[m,k]$, the number of floating-point operations times is $NN_f/H_t$. Next is the computational complexity of the reassignment operation. Brevdo et al. gave the results for wavelet-based SST [27]. For a fixed instant, if the number of frequency (or scale) bins is $n_a$, the computational complexity is $O(n_a)$. Similarly, for the proposed downsampled SST combined with selective reassignment, provided that the number of frequency bins is $zN_r$ between $f_1$ and $f_2$, the computational complexity is $O(zN_r)$ for each time step. Therefore, the computational complexity for selective reassignment and frequency subdivision is $O(zNN_r/H_t)$ and the overall computational complexity is $O(NN_f \log_2 N_f/H_t)$ considering $zN_r < N_f \log_2 N_f$ in most conditions. To sum up, the fast implementation of SST is realized due to the downsampling strategy and selective reassignment.

The computational complexity of SST reconstruction is also analyzed. The number of floating-point operations is $NN_f \log_2 N_f/H_t$ in Eq. (25) and the overlap-add method in Eq. (26) takes $LN/H_t$ times, so the overall computational complexity for inverse SST is

$$O(NN_f \log_2 N_f/H_t + LN/H_t) = O(N/H_t \cdot N_f \log_2 N_f)$$

which is the same as forward transform. For the frequency subdivision case, the computational complexity for inverse SST is multiplied by $z$.

It is noteworthy that the selective reassignment in the previous section can also reduce the computational burden for extracting TF ridges. In addition, for reconstruction with selective reassignment, one can construct $T_x$ matrix by zero paddings at $[0, f_1]$ before inverse FFT. Further, for reconstruction with frequency subdivision, one can use $zN_f$-point inverse Fourier transform after zero paddings. The rest of the steps remain the same as Eq. (11)-(26).

## 3. Parameters Selection with Numerical Study

This section presents the effects of parameters on the efficiency and accuracy of downsampled SST. To be more specific, the downsampling factor may have an impact on the computational efficiency, TF concentration, and reconstruction accuracy; the selective reassignment accelerates the computational speed; TF bandwidth mainly affects the reconstruction accuracy. Firstly, the effects on efficiency are analyzed, followed by the effects on TF concentration and reconstruction accuracy. The simulation is tested on PC with Intel(R) Core(TM) i5-8250 CPU @ 1.60 GHz, 8GB of DDR4 RAM, Windows 10 OS and MATLAB R2017b, 64bit.

The simulated signal consists of three AM-FM components $x(t) = x_1(t) + x_2(t) + x_3(t)$, where

$$x_1(t) = (1 - 0.1\cos(0.25\pi t))\cos(100\pi t)$$

$$x_2(t) = \begin{cases} \cos(500\pi t - 25\pi t^2), & 0 \leq t < 4, \\ \cos(500\pi t - 50\pi t^2 + \frac{25}{6}\pi t^3 + \frac{4}{3}\pi), & 4 \leq t < 8; \end{cases} \quad (30)$$

$$x_3(t) = (1 - 0.2\cos(0.125\pi t))\cos(740\pi t + \frac{400}{3}\sin(0.75\pi t) - 200\sin(0.5\pi t))$$

The theoretical IF for each component is calculated below and shown in Fig. 1.

$$f_1(t) = 50$$
$$f_2(t) = \begin{cases} 250 - 25t, & 0 \leq t < 4, \\ 250 - 50t + 6.25t^2, & 4 \leq t < 8; \end{cases} \quad (31)$$
$$f_3(t) = 370 + 50\cos(0.75\pi t) - 50\cos(0.5\pi t)$$

The sampling frequency $f_s$ is 1024 Hz and the number of samples $N$ is 8192. Then we compute the downsampled STFT using truncated Gaussian window with σ = 0.03s and vary $H_t$, $H_f$ and $f_w$ to see their effects on efficiency and accuracy of TFRs.

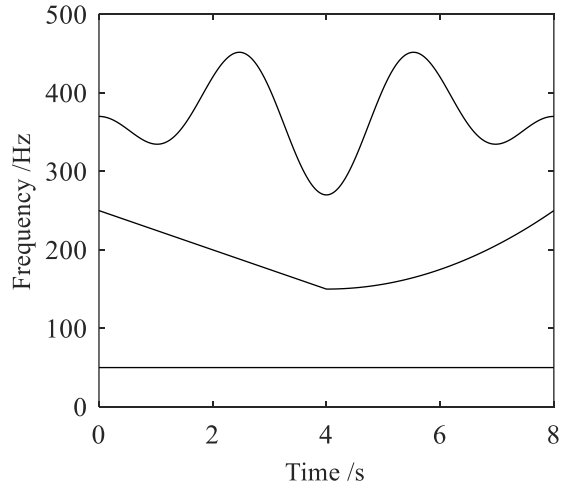

Fig. 1 IFs of the three components of the simulated signal

3.1 *Effect of downsampling factor and selective reassignment on efficiency*

The computational complexity analysis in Section 2.5 indicates that the downsampling factor and selective reassignment have a huge impact on the time cost. Fig. 2 shows the TFR of the simulated signal (SNR=5dB) with different downsampling factor, where Fig. 2(a) is a full sampled SST. Table 1 gives the elapsed time of the SST in Fig. 2. From the visual point of view, increasing $H_t$ and $H_f$ properly have little effect on the concentration or readability of TFR, but greatly improves the computational efficiency. In Table 1, the time cost in the second row is about half of that in the first row, because the number of DFT point in the second row is halved; the time cost in the second and third column is about as 1/20 and 1/40 times as that in the first column, respectively, because the number of columns of TF matrix increases decreases accordingly.

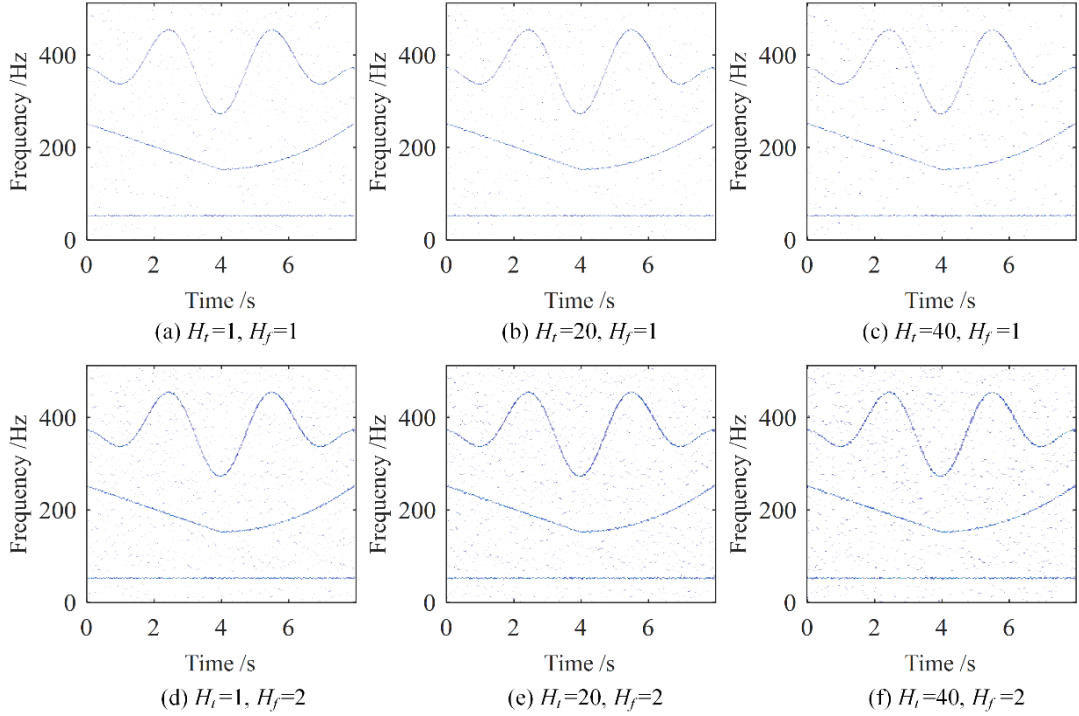

Fig. 2 TFRs of SST with different downsampling factors

Table 1 Elapsed time of SST with different downsampling factors /s (after 20 averages)

| $H_f$ | $H_t$ | | |
| --- | --- | --- | --- |
|  | 1 | 20 | 40 |
| 1 | 10.29 | 0.50 | 0.24 |
| 2 | 5.37 | 0.27 | 0.13 |

In SST, the time cost of selective reassignment should be also considered when evaluating the elapsed time. The effect of $N_r$ in the sense of big $O$ is ignored because $N_r \leq N_f$, but in practice, the reassignment operation takes much time due to its nonlinear property. Therefore, it is also important to reduce the number of frequency bins for faster reassignment. Suppose that $x_1(t)$ is the signal of interest whose frequency is 50Hz, so the frequency band to reassign is set to [0, 80] Hz with $H_f$=20. Table 2 is the result of introducing selective reassignment. In both conditions, the time cost of reassigning a subsection of the frequency band is much less than the conventional process.

Table 2  Elapsed time of SST with/without selective reassignment /s (after 20 averages)

| $H_f$ | full frequency reassignment | selective reassignment |
|---|---|---|
| 1 | 0.36 | 0.12 |
| 2 | 0.20 | 0.07 |

3.2 *Effect of downsampling factor on concentration*

As stated in the previous section, the downsampling factor, if not large enough, has little impact on the quality of TFR visually but boosts SST greatly. However, how to quantitively evaluate the effect of downsampling factor on the concentration of TFR, especially with relatively large factor, is worth studying. In this paper, Rényi entropy [28, 29] is utilized to evaluate the concentration of the TFR obtained by full sampled SST, downsampled SST and STFT. The Rényi entropy of order $\alpha$ for a TFR $P(u,\xi)$ is defined as

$$R_P^\alpha = \frac{1}{1-\alpha} \log_2 \iint \left( P(u,\xi) / \iint P(u,\xi) \mathrm{d}u\,\mathrm{d}\xi \right)^\alpha \mathrm{d}u\,\mathrm{d}\xi \tag{32}$$

The third-order Rényi entropy is used here and it decreases with the improvement of concentration.

Fig. 3 illustrates the Renyi entropy of SST and STFT with $H_t \in [1,60]$ and $H_f \in [1,100]$, where classical full sampled SST is included as $H_t=H_f=1$. Overall, SST has lower entropy under small downsampling rate, which means better TF concentration than STFT. For STFT, the entropy is rarely affected by downsampling. However, for SST, downsampling factors defined in the time domain and frequency domain have a different impact on the concentration. To be specific, the entropy of SST remains constant with $H_t$ while it increases with $H_f$ significantly. The $H_f$ has more effect on the concentration of SST since the TF energy is reassigned along the frequency axis. This coarse frequency mesh may result in relocating the coefficients which should have been relocated to different bins to the same frequency bin. Fig. 4 shows the downsampled STFT and SST with $H_f=70$. It is seen that they have similar performance, so in this case, it is unnecessary to perform SST.

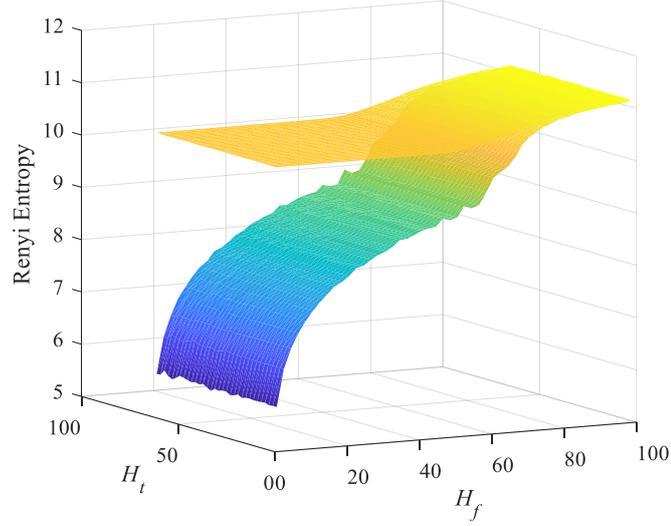

Fig. 3 TF concentration of SST and STFT at different downsampling factors

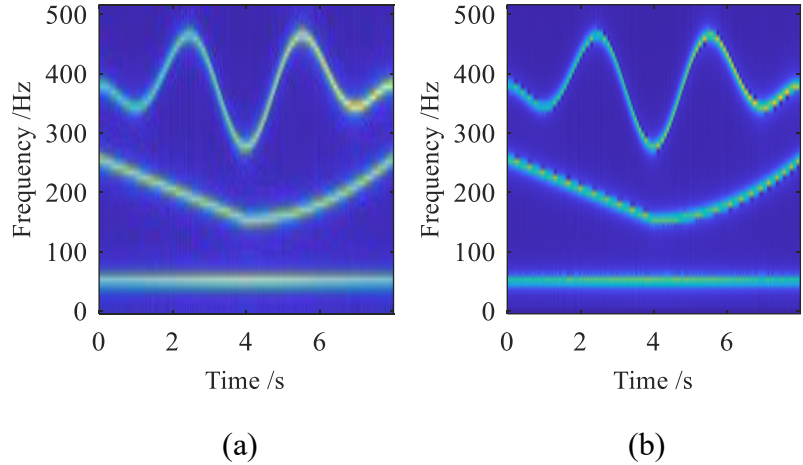

(a)                                 (b)

Fig. 4 (a) downsampled STFT and (b) downsampled SST with $H_t$=1, $H_f$=70

3.3 *Effect of downsampling factor and bandwidth on reconstruction*

    In the previous subsection, the effect of downsampling factor on concentration is studied. As one of the advantages of SST is invertible property, in this subsection, the impact of these parameters on reconstruction is studied. Moreover, compared with STFT, SST is very concentrated, so the selection of reconstruction bandwidth is not so crucial [1], which is also studied here. The reconstruction performance is evaluated by the output SNR which is defined as

$$SNR = 10\log\left\{\|x(t)\|^2 / \|x(t) - \hat{x}(t)\|^2\right\} \tag{33}$$

where $x(t)$ and $\hat{x}(t)$ denote the noise-free signal and reconstructed signal, respectively.

### 3.3.1 Effect of downsampling factor on reconstruction accuracy

The downsampled SST has both desirable invertible property and efficiency. The input SNR of the signal is 5dB and $H_t$, $H_f \in [2,4,8,16,32,40]$. Fig. 5 shows each reconstructed component of the signal (1.5~2s) with $H_f = H_t = 8$, where red is the reconstructed signal and blue is the original noiseless signal. Since Gaussian white noise added to the signal is distributed in the entire time-frequency plane, the smoothing effect of the window function causes an error on reconstructed amplitude. However, the oscillation behavior of the signal can be accurately reconstructed.

Fig. 6 shows the output SNR of downsampled SST and STFT at different downsampling factors. Overall, $H_t$ has nearly no effect on reconstruction accuracy while $H_f$ has relatively more impact but still a little. If the frequency axis is intensively sampled, SST behavior better than STFT but on a larger interval, STFT outperforms slightly SST. For comparison, full sampled SST is calculated and reconstructed using conventional formula Eq. (15), where the output SNR for $x_1$, $x_2$, and $x_3$ are 14.98, 14.57 and 12.36, respectively. Compared with full sampled SST, components reconstructed from downsampled SST using the proposed formula have similar output SNR but faster speed.

Considering output SNR among three components at fixed parameters in Fig. 6, the reconstruction accuracy of the first component is the highest while that of the third component is the lowest. This is because the first component is a harmonic signal and the SST has the strongest representation ability while the third component is the fast-varying signal which causes reconstruction error using SST.

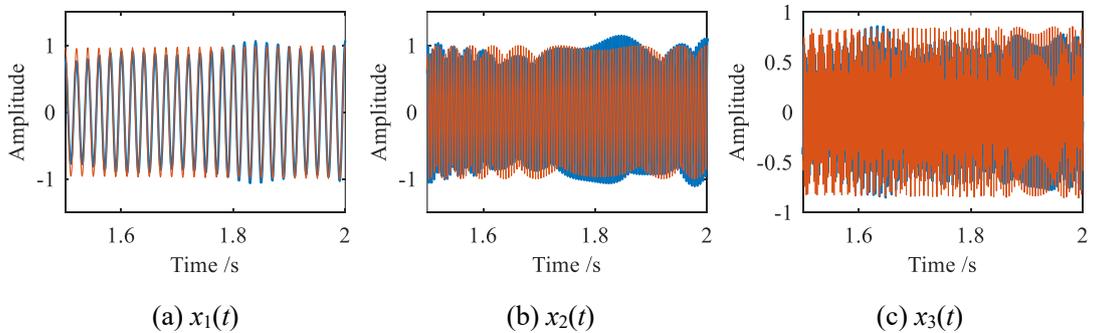

(a) $x_1(t)$      (b) $x_2(t)$      (c) $x_3(t)$

Fig. 5 Time waveform of the original signal (red) and reconstructed signal (blue), $H_f = H_t = 8$

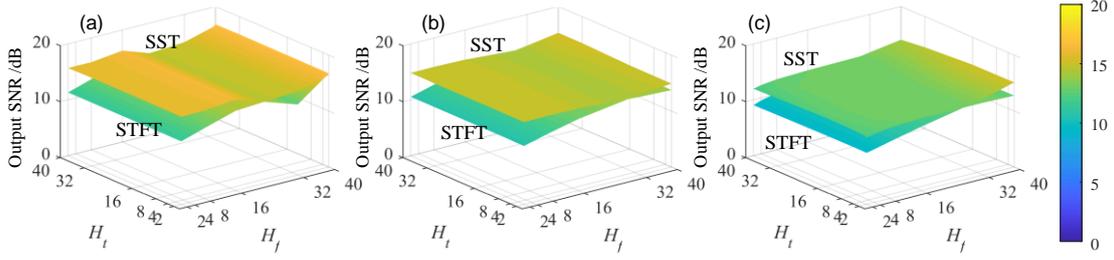

Fig. 6 Output SNR of downsampled SST and STFT of three components at different downsampling factors; (a) $x_1$, (b) $x_2$, (c) $x_3$

3.3.2 Effect of bandwidth on reconstruction accuracy

The reconstruction accuracy of SST is not sensitive to the selection of bandwidth, compared with STFT. This is validated by using different bandwidth to filter out the mode of interest in the TF plane and testing the output SNR, with input SNR=0 dB and 5 dB. Fig. 7 shows the corresponding SNR of the reconstructed signal with respect to reconstruction half bandwidth $f_w$ =0~30Hz ($H_t=H_f$ =4).

For the same input SNR, downsampled SST using the modified inverse formula not only greatly improves efficiency but also even performs better than the full SST. Another finding is that, in the downsampled framework, SST performs better than STFT in most cases, especially with small bandwidth, which indicates that one can use fewer TF points to recover signals more accurately and sparsely. The reconstruction accuracy of SST is insensitive and stable to the selection of bandwidth and one can casually choose the bandwidth for mode retrieval thanks to the robustness [27] and concentration of SST. On the other hand, mode separation by STFT is sensitive to bandwidth due to low concentration. In this case, for STFT, a relatively wide band must be carefully selected to incorporate enough coefficients for reconstruction, but a wider filter band may also introduce noise, which deteriorates the accuracy.

For each curve, the output SNR increases with $f_w$ at the beginning but falls later. This is because if the bandwidth is very narrow, the useful coefficients are lost by the narrow TF filtering, resulting in low accuracy. As the bandwidth increases, the useful coefficients are gradually incorporated and the SNR increases. However, when the bandwidth enlarges more, the noise will be also included in the TF filter, so the SNR will decrease. When the bandwidth is large enough, the STFT and full SST perform the same while the SST using the proposed inverse formula outperforms the others.

In addition, reconstruction performance of SST for a slow-varying signal like $x_1$ or $x_2$ is better than fast-vary components like $x_3$. This is an inherent characteristic of SST and is consistent with the result in Section 3.3.1. On the contrary, STFT has the same performance over different components.

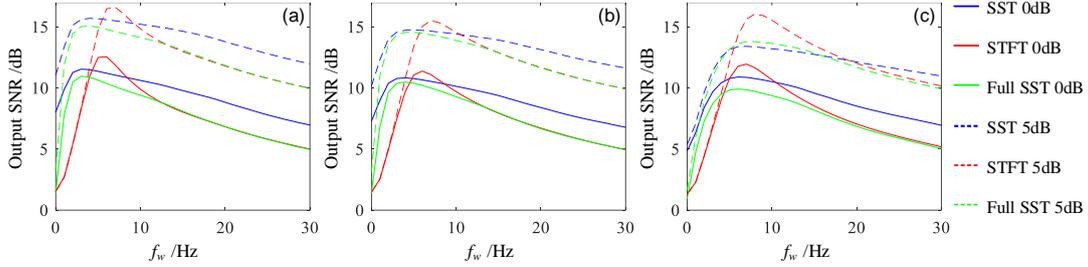

Fig. 7 Reconstruction accuracy of SST, STFT, and full SST at different bandwidth and input SNR
(a) $x_1$ (b) $x_2$ (c) $x_3$

## 4. Reassignment behavior with downsampling factors

In Section 3.2, the effects of downsampling factors on TF concentration are presented via simulation. In order to generalize this simulation and analyze when *exactly* the downsampled SST and STFT have similar concentration, a mathematical model of reassignment behavior with downsampling factors is studied. The downsampling factors here are $H_f$ and $H_t$. Since for SST, only the $H_f$ affects reassignment behavior so $H_t$ is not considered.

Firstly, the theoretical frequency diffusion of a TF ridge is obtained, based on Eq. (18). At a fixed time instant $u$, the frequency representation of Gaussian window is

$$\hat{g}(\omega) = (\pi\sigma^2)^{-1/4} e^{-\sigma^2\omega^2/2} = (\pi\sigma^2)^{-1/4} e^{-\frac{\omega^2}{2\cdot(1/\sigma)^2}} \tag{34}$$

This means that the coefficients in the TF plane are concentrated around so-called ridges, defined by $\xi = f_k(u)$. Thus, the theoretical frequency diffusion can be represented by frequency standard deviation

$$\sigma_f = \frac{\sigma_\omega}{2\pi} = \frac{1}{2\pi\sigma} \tag{35}$$

In the simulation of the previous section, $\sigma_f = 5.31\,\text{Hz}$. The effective support in frequency is about $6\sigma_f = 31.8\,\text{Hz}$.

The TF concentration changes only when the coefficients are relocated on the frequency axis. The occurrence of reassignment depends on the IF estimator in Eq. (13) and the frequency interval $\Delta f$. For a fixed time $u$, when $|\hat{\omega}_x[k] - \xi[k]|$ in a frequency bin is larger than the half interval $\Delta f/2$, the coefficient $S_x^g[k]$ in this position $k$ will jump to the adjacent or farther frequency bins according to the distance $|\hat{\omega}_x[k] - \xi[k]|$. Take the 50Hz harmonic component in Fig. 4 for example. The half frequency interval is about 4.5Hz ($H_t$=1, $H_f$=70) and the discrete STFT magnitude distribution around 50Hz is shown in Fig. 8. The green numbers are the signed distance to move for each bin in the effective support [50-3$\sigma_f$, 50+3$\sigma_f$]. However, they are all smaller than half frequency interval, so TF coefficients in these four bins cannot jump to other bins. Therefore, the concentration or entropy does not change, as illustrated in Fig. 4.

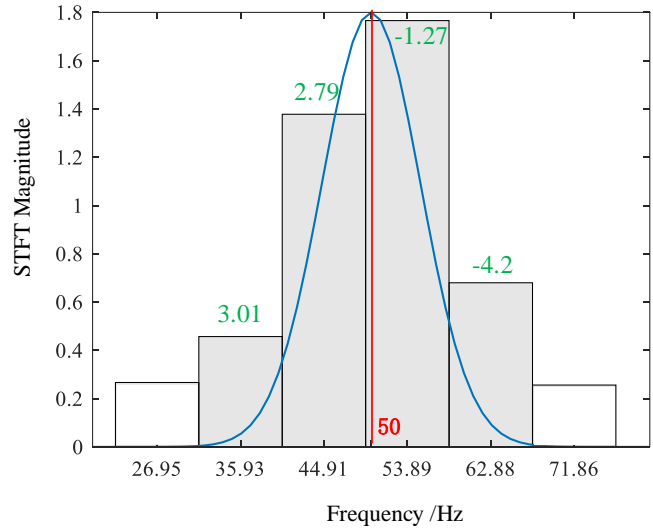

Fig. 8 The discrete STFT magnitude distribution and RF (green numbers) in each frequency bin

The farther a TF point is from the theoretical IF, the larger the distance to move will be and the more likely this point will move. As long as it moves, the concentration will increase. The distance for a TF point to move is defined as reassignment frequency (RF)

$$\omega_r(u,\xi) = \hat{\omega}_x(u,\xi) - \xi = -\Im\left[S_x^{g'}(u,\xi)/S_x^g(u,\xi)\right] \qquad (36)$$

It is negative when a point is located at a higher frequency than theoretical frequency and vice versa. Considering the $\omega_r(u,\xi)$ on a slice at 1.3s, for example, the length of the linear part is about 30Hz in Fig. 9(a), which is corresponding to the effective support obtained in Eq. (35). When the $H_f$ increases, the $|\omega_r(u,\xi)|$ decrease (see Fig. 9(b)) and it is less likely for a point to move. Thus, the moment when the entropy of SST does not decrease is enlarging $\Delta f$ until the farthest point in effective support does not move. This is shown in Fig. 10, where in the effective support of $\hat{g}(\omega)$ there are four frequency bins and the two central ones (gray) which incorporate the theoretical IF dominate the energy. When the coefficients in the two white bins outside the two dominated bins do not move, the maximum interval which allows reassignment is obtained. i.e.

$$\Delta f < \frac{3}{2}\sigma_f \tag{37}$$

Combined with Eq. (12) and Eq. (35) we have

$$H_f < \frac{3N}{4\pi\sigma f_s} \tag{38}$$

so the upper limit of parameter $H_f$ can be approximated by demonstration; for this simulation, $H_f<64$.

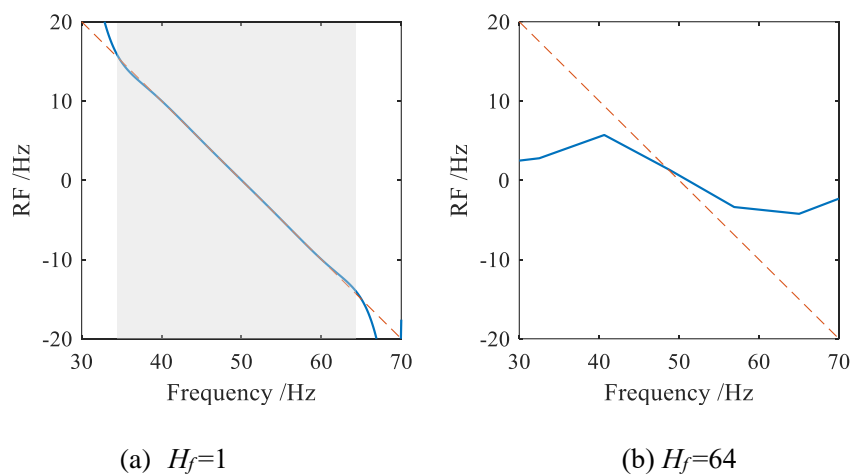

(a) $H_f=1$    (b) $H_f=64$

Fig. 9 The absolute value of RF decreases with $H_f$

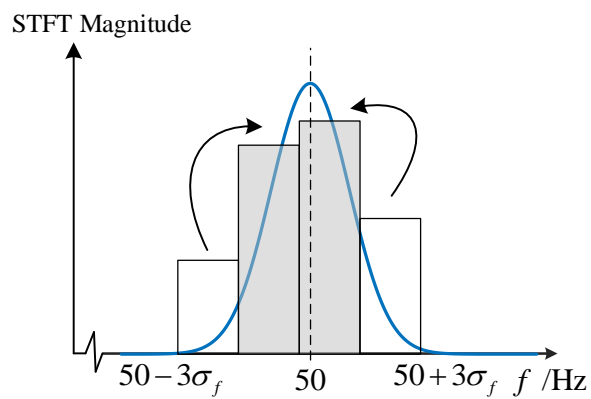

Fig. 10 The reassignment behaviour of coefficients at critical frequency interval

In order to formulize the effect of frequency discretization on the reassignment process, for Gaussian window, the following holds

$$S_x^{g'}(u,\xi) = \int x(t)g'(t-u)e^{-i\xi(t-u)}\,dt = \frac{-1}{\sigma^2}S_x^{tg}(u,\xi)$$

$$\partial_\xi S_x^g(u,\xi) = -i\int x(t)(t-u)g(t-u)e^{-i\xi(t-u)}\,dt = -iS_x^{tg}(u,\xi) \quad (39)$$

$$\Im\left\{\frac{S_x^{g'}(u,\xi)}{S_x^g(u,\xi)}\right\} = -\frac{1}{\sigma^2}\frac{\partial_\xi S_x^g(u,\xi)}{S_x^g(u,\xi)}$$

For slow-varying signals suitable for SST, combined with Eq. (18), (34) and (36), we get

$$\omega_r(u,\xi) = \frac{1}{\sigma^2}\frac{\partial_\xi S_x^g(u,\xi)}{S_x^g(u,\xi)} \approx \frac{1}{\sigma^2}\frac{\partial_\xi \hat{g}(\xi-f_x(u))}{\hat{g}(\xi-f_x(u))}$$

$$= \frac{1}{\sigma^2}\frac{\partial_\xi e^{-(\xi-f_x(u))^2\sigma^2/2}}{e^{-(\xi-f_x(u))^2\sigma^2/2}} \quad (40)$$

Let $\xi_d(u) = \xi - f_x(u)$ which means the signed distance from the ridge and we have

$$\omega_r(u,\xi) \approx \frac{1}{\sigma^2}\frac{\partial_\xi e^{-\xi_d(u)^2\sigma^2/2}}{e^{-\xi_d(u)^2\sigma^2/2}} \quad (41)$$

This means $\omega_r(u,\xi)$ can be calculated by derivative with respect to $\xi_d(u)$ in the continuous form. However, in discrete form, the difference formula introduces errors, especially when $\Delta\xi$ is relatively large. For simplicity, for a fixed time $u$, Eq. (41) can be rewritten as

$$\omega_r(\xi) \approx \frac{1}{\sigma^2}\frac{\partial_\xi e^{-\xi_d^2\sigma^2/2}}{e^{-\xi_d^2\sigma^2/2}} \quad (42)$$

with its discrete version

$$\hat{\omega}_r \approx \frac{1}{\sigma^2}\frac{e^{-(\xi_d+\Delta\xi)^2\sigma^2/2} - e^{-\xi_d^2\sigma^2/2}}{e^{-\xi_d^2\sigma^2/2}\Delta\xi}$$

$$= \frac{1}{\Delta\xi\sigma^2}\left(e^{-\Delta\xi^2\sigma^2/2 - \xi_d\Delta\xi\sigma^2} - 1\right) \quad (43)$$

Let $\Delta f = \Delta\xi/2\pi$ and $f_d = \xi_d/2\pi$, so the RF is approximated as

$$\hat{f}_r = \frac{\hat{\omega}_r}{2\pi} \approx \frac{1}{4\pi^2\Delta f\sigma^2}\left(e^{-2\pi^2\Delta f^2\sigma^2 - 4\pi^2\Delta f\sigma^2 f_d} - 1\right) \quad (44)$$

when $\Delta f \to 0$, $\hat{f}_r \to -f_d$, i.e. the $|\hat{f}_r|$ increases linearly with $|f_d|$, but when $\Delta f$ cannot be neglected, $|\hat{f}_r|$ deviates from linearity, as shown in Fig. 11. To prevent the global reassignment, the TF points as far as possible from the frequency center should not be relocated, i.e.

$$\left|\hat{f}_r\right| < \left|\lim_{|f_d| \to \infty} \frac{1}{4\pi^2 \Delta f \sigma^2}\left(e^{-2\pi^2 \Delta f^2 \sigma^2 - 4\pi^2 \Delta f \sigma^2 f_d} - 1\right)\right| = \frac{1}{4\pi^2 \Delta f \sigma^2} < \frac{\Delta f}{2} \tag{45}$$

This is the condition when entropy does not decrease, so for effective SST, one should have

$$\frac{\Delta f}{2} < \frac{1}{4\pi^2 \sigma^2 \Delta f}$$

and

$$\Delta f < \sqrt{2}\sigma_f \tag{46}$$

which is an alternative method for approximation of $H_f$ and it is close to Eq. (37). In this simulation, $H_f < 60$. Both Eq. (37) and (46) can provide the theoretical guideline for parameter selection on the downsampling factor when using SST in the practical engineering field.

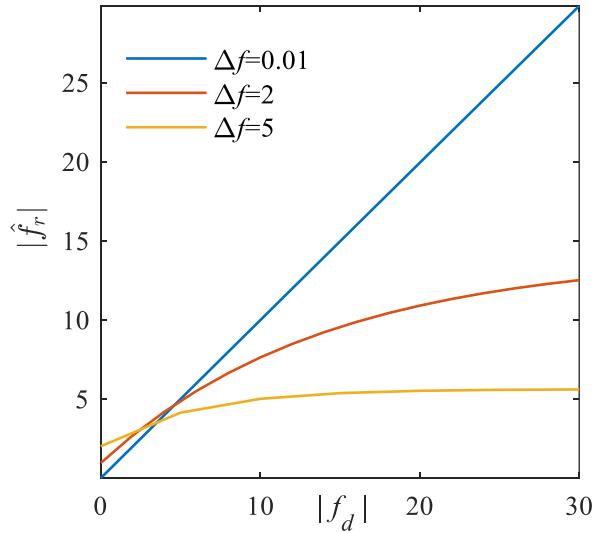

Fig. 11 The estimated RF causes error at coarse frequency interval

## 5. Experimental Validation

### 5.1 *Vibration signal analysis from aero-engine*

The effectiveness of the proposed downsampled SST is validated by vibration data from the aircraft engine on Surveillance 8 challenge [30]. The data is acquired by an accelerometer mounted on the intermediate case near the radial drive shaft with a sampling rate 44.1 kHz and a duration of 200s. The number of samples is $N$=9035089, so such large-scale data cannot be processed by standard full SST. In this part, downsampled SST is employed to characterize high-frequency speed fluctuations of L4 shaft from vibration signal and reconstruct the dominant components for post-processing.

The analysis parameters are set as follows: the number of FFT points is $N_f=2^{13}$ and the downsampling factor on frequency $H_f=N/N_f\approx1103$, less than the critical value 2446 calculated by Eq. (38). The overlapping ratio of adjacent windows is 80%, which means $H_t = 0.2L = 1149$. In order to further boost computing efficiency without losing IF estimation results, selective reassignment and frequency subdivision are both employed, where the frequency range is [6900, 9500] and subdivision factor $z$=8. Fig. 12 shows the dominant TF curve in STFT and SST with σ=0.02 and their enlarged view. The SST in Fig. 12(c) obtains better TF concentration and characterizes fluctuating IF better with $\Delta f = 0.67 \text{Hz}$ while the downsampled STFT in Fig. 12(b) can only depict the overall trend. The elapsed time of downsampled SST is only about 3.9s after 20 averages.

To further illustrate the superior performance of the proposed method, TF ridges are extracted from STFT and SST on 100~200s to compare the extracted IF with the one calculated by the tachometer. The tachometer is mounted on L4 shaft and has a resolution of 44 pulses per revolution. The L4 shaft speed can be obtained by the phase demodulation method [31]. Then the extracted IF from TFR is normalized to the speed of L4 shaft to be comparable. The comparison between the extracted IF from TFRs and the one calculated by the tachometer is shown in Fig. 13. The IF obtained by SST agrees well with the IF detected by the tachometer while the STFT produces coarse estimation results. The results indicate that the proposed method can characterize weak speed fluctuations without carefully selecting filter band or smooth operation as phase demodulation does because TFR itself has an inherent smooth property.

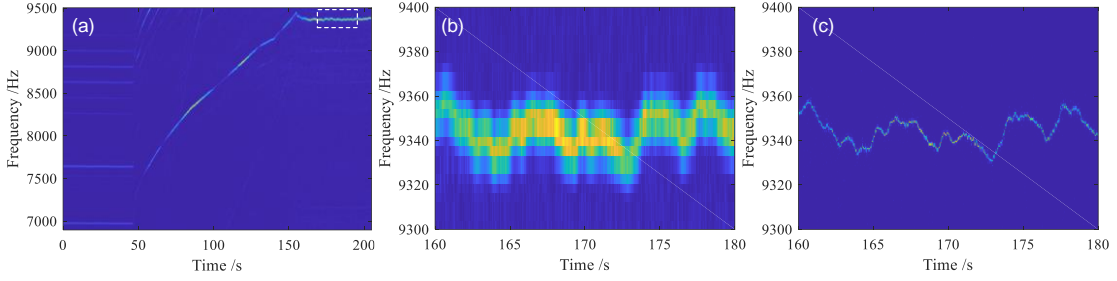

Fig. 12 The TFRs of the vibration signal: (a) downsampled STFT; (b) the enlarged view of the dash line box in (a); (c) the enlarge view of downsampled SST.

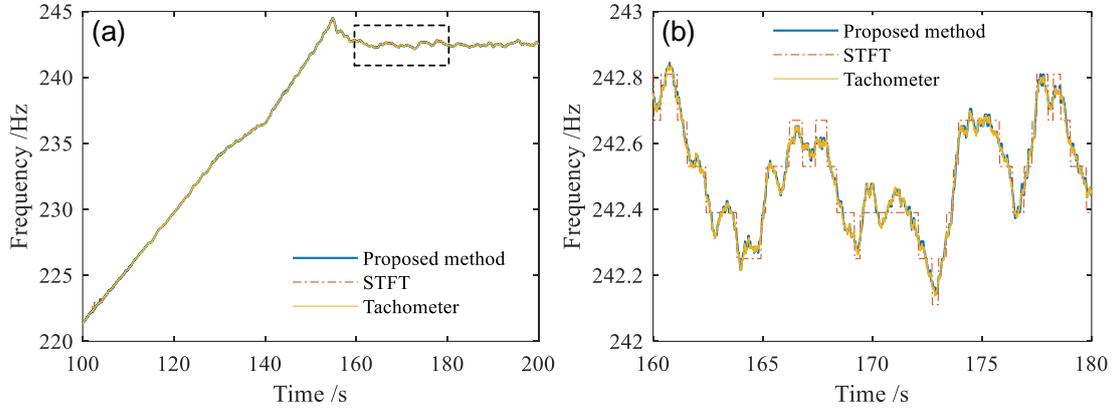

Fig. 13 (a) IF estimated by different methods (100~200s) and (b) the zoomed view (160~180s)

In this study, the reconstruction property of downsampled SST is exploited by modes recovery from the collected acceleration signal. The half bandwidth $f_w$ is 20 and 10 for STFT and SST, respectively. Fig. 14 shows the reconstructed mode from proposed SST, which coincides well with the signal reconstructed by STFT. The error between these two signals is measured by root-mean-square error (*RMSE*)

$$RMSE = \sqrt{\frac{1}{n}\sum_{i=1}^{n}(\hat{x}(i)-x(i))^2} \qquad (47)$$

where $\hat{x}(i)$ is the reconstructed signal and $x(i)$ is the reference signal. In this case, the *RMSE* value is about 0.0052, which indicates that the proposed downsampled SST method can approximately but directly reconstruct signal, using smaller bandwidth compared with STFT.

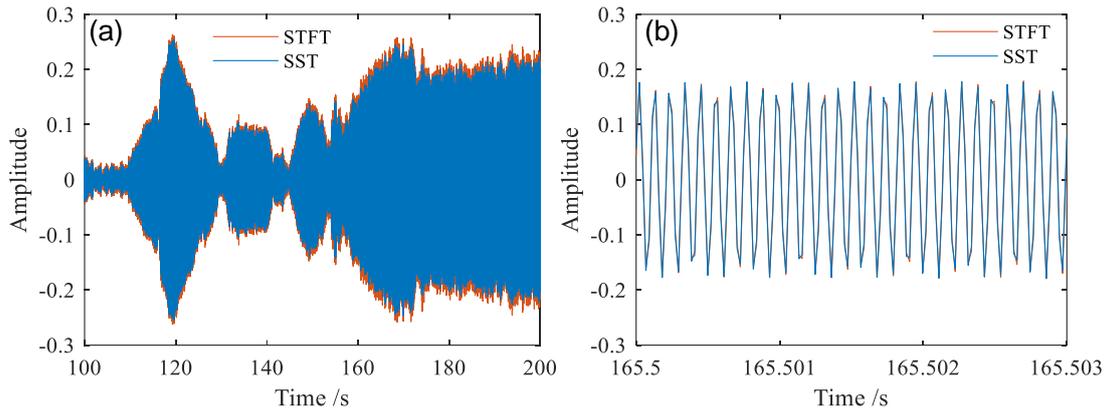

Fig. 14 (a) Reconstructed components from STFT and SST with (b) enlarged view

### 5.2 *Chatter identification*

Vibration signals on chatter conditions have typically modulated and nonlinear features, so they need to be analyzed by TFR methods rather than classical spectrum analysis. In this subsection, the proposed SST is thus employed to investigate the non-stationary feature of vibration signals under chatter conditions. In the cutting experiment, a trapezoid workpiece is employed to vary the cutting stable conditions from left to right along with increasing cutting depth. The nominal speed of the spindle is 9600 r/min and vibration signals are collected by a PCB accelerometer in X and Y direction, as shown in Fig. 15(a).

Fig. 16 shows the time-vary features extracted from TFRs of vibrational signal with a sampling rate of 10240 Hz and a duration of 56s. The number of samples is $N$=573440, so it is difficult for a PC to process such large-scale data using standard SST. Fig. 16(a) is the time waveform of the acceleration signal. The chatter frequency usually exists near the lower natural frequency of the spindle-toolholder-tool (STT) system, so the frequency range for selective reassignment is set 1200~1400Hz to search chatter frequency more efficiently according to the frequency response function (FRF) of the STT system in Fig. 16(b). The FRF is measured by an accelerometer and hammer shown in Fig. 15(b).

Fig. 16(c) and (d) show the SST and STFT results, respectively, with window scale parameter $\sigma$=0.1s, $N_f$=$2^{15}$, and $H_t$=667. For downsampled SST in Fig. 16(c), selective reassignment on 1200~1400Hz is used but without subdivision. Obviously, the SST is more concentrated than STFT and the elapse time of SST is only 0.8s after 20 averages.

The time-varying chatter frequency component starting at point A and A' can be identified clearly, caused by the modal parameters change during thinwalled milling [32]; the time instant for chatter occurrence can be determined at about 28s.

In order to investigate the relationship between the rotational speed of the spindle and chatter occurrence, the fourth harmonic of rotational frequency is analyzed by the proposed SST in Fig. 16(e), with $H_t$=200, selective reassignment at 630~640Hz and subdivision factor $z$=20. The proposed SST costs 3s after 20 averages and it can characterize the weakly but quickly fluctuating IF features of spindle speed, which cannot be seen in STFT due to lack of concentration and selective reassignment (the frequency interval for STFT is $f_s/N_f$=0.3Hz). In Fig. 16(f), the IF of the component in (e) is extracted. In stable cutting conditions, the IF fluctuates weakly and quickly before point B. However, as the axial depth of cut increases, the IF has a more significant low-frequency component and a slightly higher fluctuating amplitude, resulting in a more complex composition. This is because when the chatter occurs, the dynamic cutting thickness formed by the adjacent teeth is very different along with time, causing strong-varying cutting force that is disordered by the adjacent teeth. This leads to a more chaotic cutting excitation and a chaotically modulated vibration response. The existence of the irregular low-frequency components and the increase in fluctuating amplitude of IF of the vibration response indicate the time-varying chatter status. Thus, point B on the IF curve of the rotational frequency can be used to reveal the chatter occurrence, which is consistent with the time instant A in Fig. 16(c) and (d).

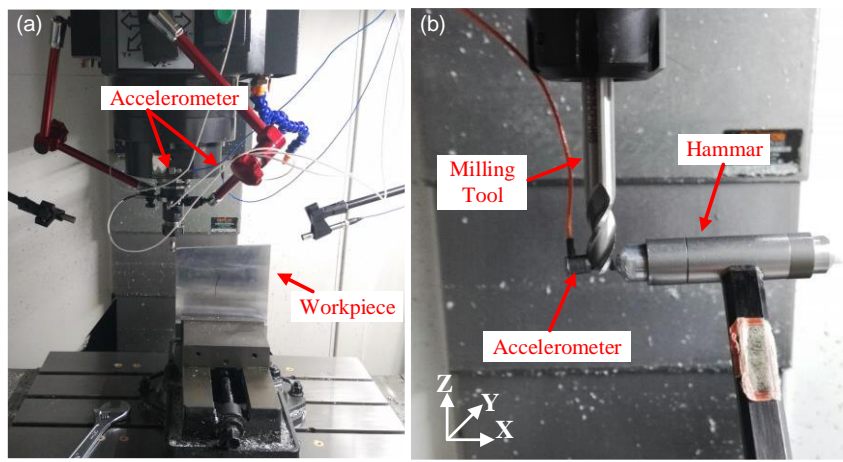

Fig. 15 The experimental setup for (a) acceleration signal collection on the milling process and (b) the FRF test of STT system

It should be noted that the maximum fluctuating frequency which can be effectively characterized by the proposed SST is limited by the time interval between adjacent columns of the TF matrix. It is further constrained by parameter $H_t$. In order to characterize higher-frequency fluctuations, one needs to reduce $H_t$, so the tradeoff between computing efficiency and the ability for characterizing fluctuations needs to be considered.

The reconstruction property of downsampled SST is also validated by such vibration data under stable and chatter conditions. The time-varying component in the interval of adjacent harmonics of rotational frequency is taken out in Fig. 16(c), from 1300 to 1400 Hz. It is then reconstructed through SST and STFT for comparison. Fig. 17 shows the reconstructed component from the proposed SST, which coincides with the signal reconstructed by STFT with $RMSE$ = 0.06. Although there are amplitude errors, the oscillation behavior of this component can be accurately reconstructed without phase shifting. This method thus provides the possibility for separating chatter-sensitive components efficiently for post-processing such as statistical indicators calculation [33].

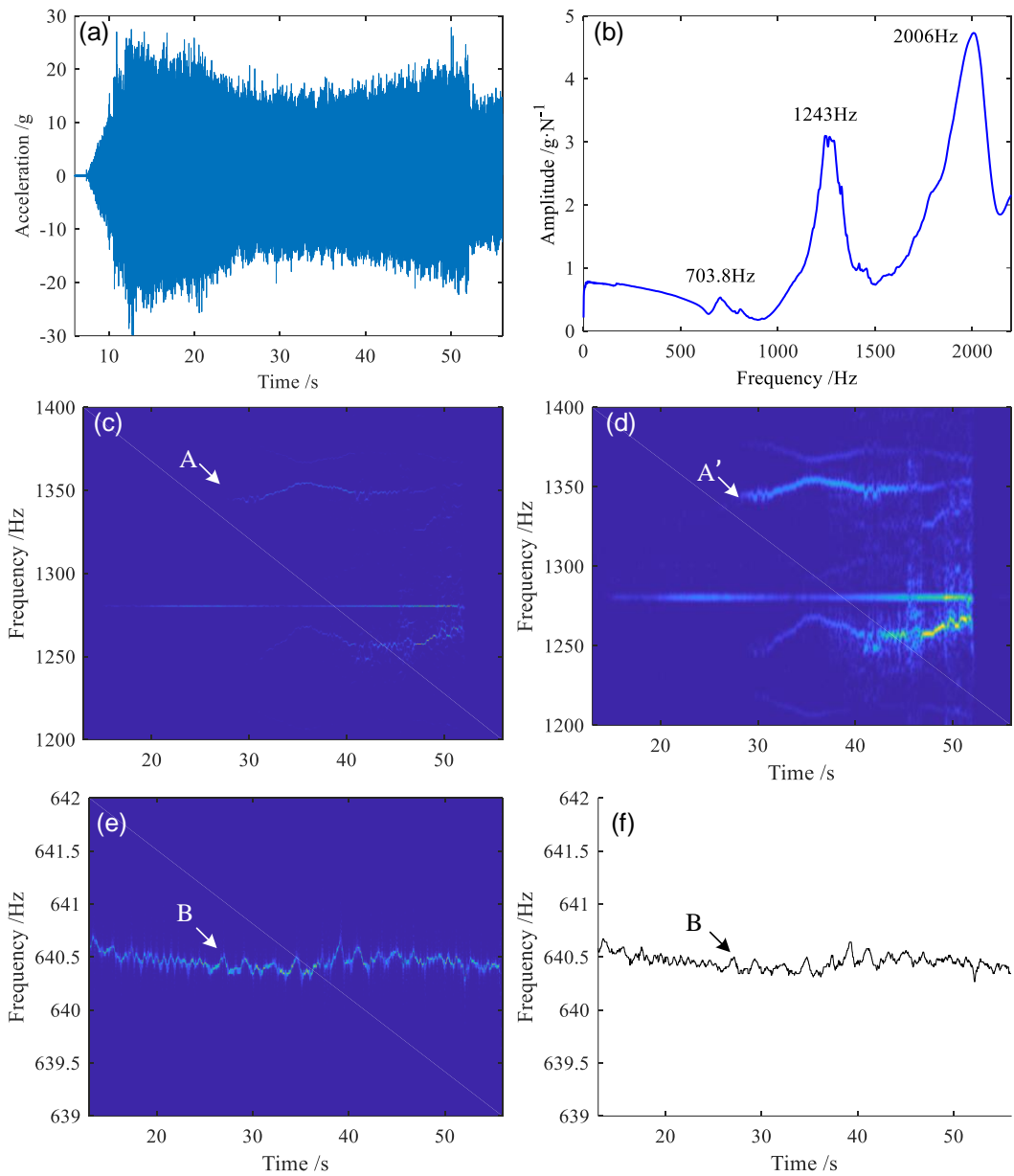

Fig. 16 Time-varying features of vibration signal in chatter condition extracted by TFR. (a) time waveform of acceleration signal; (b) FRF of the STT system in X direction; (c) SST and (d) STFT near the second resonance band of FRF; (e) SST of the fourth harmonic of rotational frequency on the spindle; (f) IF extracted from (e).

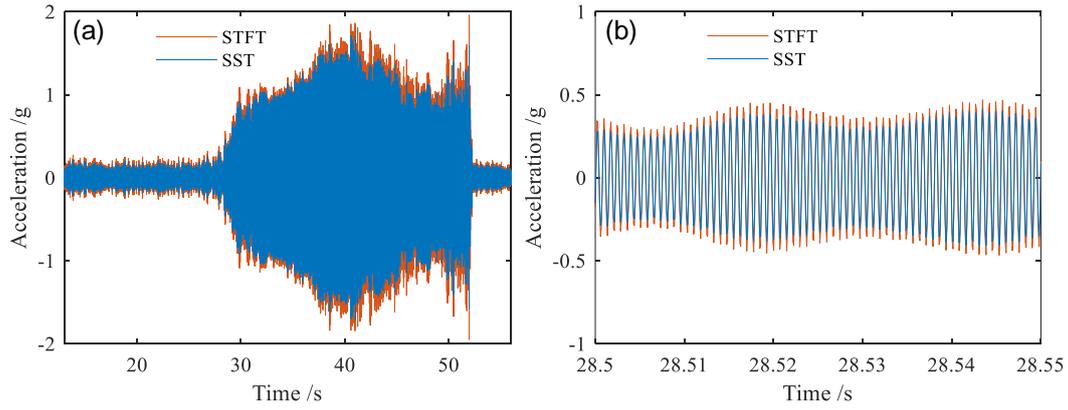

Fig. 17 The reconstructed component which is sensitive to chatter. (a) retrieved mode to illustrate cutting condition from stable to chatter; (b) the enlarged view of (a).

## 6. Conclusions

In this paper, to overcome the limitations of processing large-scale vibration signals using full SST, a fast implementation of SST based on downsampled STFT combined with selective reassignment and frequency subdivision is proposed. More importantly, the reconstruction property is also available in this fast framework accomplished by an approximate but direct inverse formula. The computational complexity analysis and simulation results show the downsampling factor and selective reassignment can reduce the computational burden. The downsampling factor on frequency has more effect on concentration than that on time. When SST is very sparsely sampled along the frequency axis ($\Delta f \approx 1.5\sigma_f$), its performance degenerates to STFT. In general, the downsampling factor has little effects on reconstruction accuracy if properly selected but one needs to consider the trade-off between the IF extraction result and efficiency. As for bandwidth, different from STFT synthesis, inverse SST is insensitive to the selection of bandwidth for mode retrieval.

Experimental results on an aero-engine and a spindle system show that the proposed fast SST can effectively characterize the non-stationary characteristics of the vibration signal. The time-varying non-stationary characteristics of the vibration response, such as IFs and separated modes of interest, can be further exploited to reveal the mechanism of self-excited chatter in the milling process, or condition monitoring the on aero-engines.

One limitation of our reconstruction formula of downsampled SST is that it is an approximation without inverse mapping needed. Fortunately, the idea of the inverse map proposed in [21] can also be inserted into our scheme to improve the accuracy in the future. Further acceleration can be accomplished by converting the calculation of STFT to filter bank-based in the frequency domain. This can be incorporated into the selective reassignment of SST since filter-based STFT itself can be recursively implemented at a predetermined frequency range. In addition, the downsampled SST and its inverse transform can be further extended to second-order for the capability of processing more strong time-varying signals according to practical needs.

## Acknowledgments

This work is supported by National Natural Science Foundation of China (Nos. 51575423, 11772244).